# Fast and energy-efficient non-volatile III-V-on-silicon photonic phase shifter based on memristors

Zhuoran Fang[1,2*], Bassem Tossoun[1], Antoine Descos[1], Di Liang[1], Xue Huang[1], Geza Kurczveil[1], Arka Majumdar[2,3], Raymond G. Beausoleil[1]

[1]*Hewlett Packard Labs, Hewlett Packard Enterprise, Milpitas, CA 95035, USA*

[2]*Department of Electrical and Computer Engineering, University of Washington, Seattle, WA 98195, USA*

[3]*Department of Physics, University of Washington, Seattle, WA 98195, USA*

*Email: rogefzr@uw.edu

## Abstract

Silicon photonics has evolved from lab research to commercial products in the past decade as it plays an increasingly crucial role in data communication for next-generation data centers and high performance computing[1]. Recently, programmable silicon photonics has also found new applications in quantum[2] and classical [3] information processing. A key component of programmable silicon photonic integrated circuits (PICs) is the phase shifter, traditionally realized via the thermo-optic or plasma dispersion effect which are weak, volatile, and power hungry. A non-volatile phase shifter can circumvent these limitations by requiring zero power to maintain the switched phases. Previously non-volatile phase modulation was achieved via phase-change[4] or ferroelectric materials[5], but the switching energy remains high (pico to nano joules) and the speed is slow (micro to milli seconds). Here, we report a non-volatile III-V-on-silicon photonic phase shifter based on $HfO_2$ memristor with sub-pJ switching energy (~400fJ), representing over an order of magnitude improvement in energy efficiency compared to the state of the art. The non-volatile phase shifter can be switched reversibly using a single 100ns pulse and exhibits an excellent endurance over 800 cycles. This technology can enable future energy-efficient programmable PICs for data centers, optical neural networks, and quantum information processing.

## Introduction

The silicon photonic phase shifter is a fundamental building block of programmable PICs[6]. For certain applications such as electro-optic modulators, the phase shifters need to operate fast and are generally achieved by free-carrier dispersion[7] or electro-optic effect[8,9], which are ultra-fast but the extent of index change is small. On the other hand, applications such as light routing between functional devices and post-fabrication trimming, do not require high speed but need large change in index. These functionalities are traditionally achieved using thermo-optic effect[10] and micro-electro-mechanical systems (MEMS)[11]. A major limitation of all these methods is their volatile nature – a constant power (~mW) or bias (>20V) must be applied to maintain the switched states. For programmable photonics, it is highly desirable to have non-volatile phase shifters which require zero[4,12] or minimal power[5] to maintain the switched states and the information is retained when the bias is removed. For examples, non-volatile phase shifters based on phase change materials[4,13] are very compact and allow true 'set-and-forget' operation, but they require relatively high switching energy (~nJ) and suffer from slow switching speed (~hundreds ns to μs). The high temperature (~900K) required to switch the phase change materials can also have reliability concerns. Non-volatile ferroelectric phase shifters based on Barium Titanate[5] can achieve multilevel operation and low loss, but the electro-optic effect is weak, and the switching speed is even slower, requiring $10^4$ pulses with a total duration of hundreds of microseconds for initializing the states. Although MEMS are shown to exhibit non-volatile effect and has compact footprint ($L_\pi$~17μm), they require large driving voltage (>20V) and the switching time can take up to a few seconds imposed by the slow I-V sweep[14] to achieve adhesion latching. Lastly, plasmonic memristor switches have been demonstrated to exhibit the latching effect[15,16], but the high insertion loss (~10dB) from the metal absorption makes it prohibitive for cascaded phase shifters. Supplementary information Table 1 presents a comparison of non-volatile programming technologies in PICs. Here, we demonstrate a non-volatile phase shifter on the III-V-on-silicon heterogeneous platform enabled by memristor effect[17–20]. The non-volatility arises from resistively switching the InP-HfO$_2$-Si memristor between high-resistance-state (HRS) and low-resistance-state (LRS) using a single 100ns long voltage pulse. The phase shifter exhibits a switching energy as low as ~400fJ, representing over an order of magnitude reduction compared to the state-of-the-art[4,5], and an excellent endurance of over 800 cycles. Moreover, the III-V-on-silicon heterogeneous platform is fully compatible with foundry processes[21], potentially enabling a seamless integration with laser sources and the energy efficient large-scale programmable PICs.

## Working principle of the III-V-HfO$_2$-Si memristor

The memristor is based on the heterogeneous integration between n-type InP and p-type silicon sandwiching a high dielectric constant HfO$_2$[19,22], see Fig. 1ai. In Fig. 1ai, the memristor is in its as-fabricated state and applying a negative bias $V_{read}$ on the n-InP while keeping the p-Si grounded (i.e., forward bias) will cause carrier accumulation at the oxide-semiconductor interface (Fig. 1bi), which essentially operates as a MOS (metal-oxide-semiconductor) capacitor. On the other hand, if a large enough positive bias is applied, oxygen vacancies (O.V.) diffuse inside the oxide layer and, at a certain threshold, a conduction filament (CF) consisting of oxygen vacancies is formed[25] (see Fig. 1aii). Such process is termed 'electroforming' and can be visualized in Fig. 1c green line where the current suddenly increases at around 7.5V and hits the current compliance enforced by the source meter. The voltage in Fig. 1c is the bias applied to the n-InP. After forming, the memristor is switched into the LRS and the MOS capacitor effect is suppressed due to the charge leakage. This is reflected in the Fig. 1bii where the forward bias no longer causes carrier accumulation. To return the memristor to HRS, a bias with opposite polarity (i.e., negative) is applied to the n-InP that causes Joule heating and the retraction of oxygen vacancies, eventually giving rise to the rupture of the CFs. The operation is called RESET and is revealed by blue lines in Fig. 1c, where the current rapidly increases to around 2.5μA at around -2V before collapsing to near zero. Since Joule heating depends on the power not on polarity, if the CF's rupture was only caused by the Joule heating, then we would have seen the CF's rupture during the voltage ramp down after the CFs form as higher power is measured (Fig. 1c orange lines). The absence of filament breaking at positive bias shows that the RESET operation is a compound phenomenon of both field effect and Joule heating that enables the switching at lower voltage than the SET[26]. Reversing the polarity causes the retraction of oxygen vacancies which facilitates the rupture of the filaments. Indeed, electrochemical migration of oxygen vacancies has been regarded as the driving mechanism behind bipolar switching where the SET and RESET polarities are opposite[27]. The rupture of the CFs brings the memristor back to the HRS and MOS capacitor effect is restored (Fig. 1aiii and 1biii). The memristor can then be switched to the LRS again by applying a positive bias to the n-InP but at lower voltage (5-6V), as shown by the orange lines in Fig. 1c. This is because the CFs do not rupture completely during RESET (Fig. 1aiii) and less oxygen vacancies need to be displaced by the field to re-connect the CFs. Fig. 1c shows that SET and RESET operations are repeated for seven consecutive cycles between HRS and LRS.

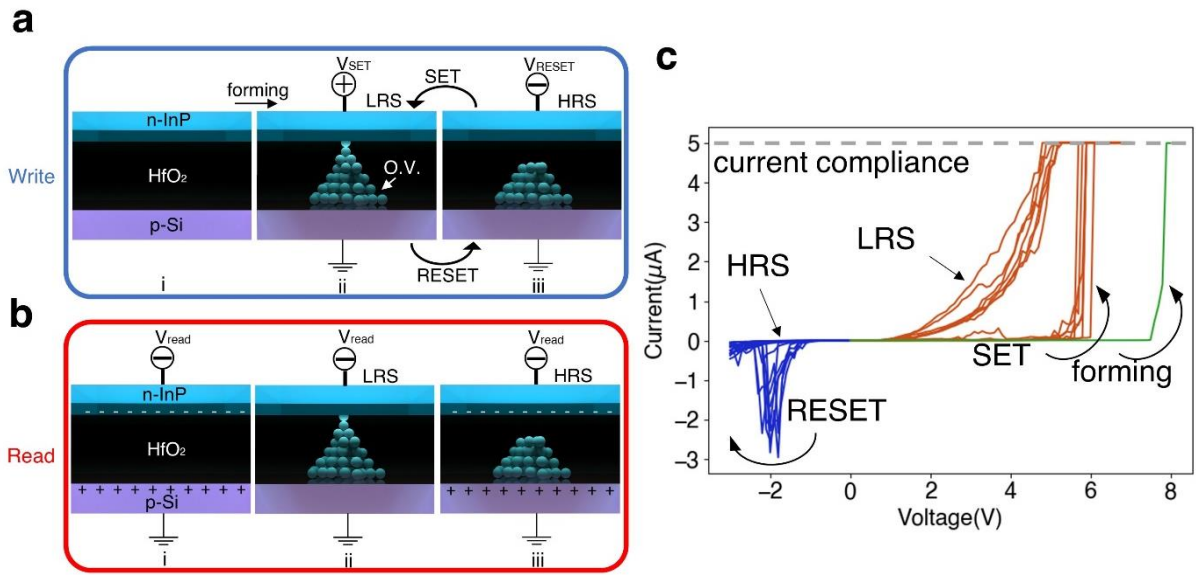

**Figure 1: Write and read states in a III-V-HfO$_2$-Si memristor.**

**a** Schematics illustrating how the states in InP-HfO$_2$-Si memristor is written. HRS (LRS), high resistance state (low resistance state). O.V. (oxygen vacancies). **b** Schematics showing reading the InP-HfO$_2$-Si memristor at states shown in **a**. A constant negative read bias is applied to the n-InP in i, ii, and iii. **c** Current voltage relationship of the forming (green), SET (orange), and RESET (blue) operations. The voltage is the write bias. The grey dash line indicates the 5μA current compliance to prevent device damage. Seven I-V SET and RESET cycles are plotted on top of each other to show good cyclability.

## A memristor-based III-V-on-silicon heterogeneous photonic platform

The ability to switch a MOS capacitor on and off using a memristor provides a way to realize non-volatile phase tuning in a microring resonator by controlling the carriers in the accumulation layer formed at the waveguide (Fig. 2a). Thus, the optical modulation happens via the free-carrier dispersion effect, while the non-volatility comes from the memristor controlling the MOS capacitor, which in turn controls the ability to form an accumulation layer of free carriers. The 250nm tall, 1.2μm wide silicon waveguide is formed by partially etching to a 100nm slab on one side and fully etched to the buried oxide (BOX) on the other side. A 150nm n-InP epitaxial layer is transferred to the silicon layer via standard wafer bonding technique[23] with 9.6nm-thick HfO$_2$ as the interface oxide[22] (see Methods for fabrication details). The HfO$_2$ acts simultaneously as a gate oxide that forms a MOS capacitor with the Si and InP, and a memristor switching layer. The air gap is left in the Si layer to confine the MOS capacitor to the fundamental TE mode area in the Si waveguide and hence reduce the total capacitance for a fast electro-optic response[24]. Fig. 2b shows the fabricated microring resonator integrated with the memristor. In the as-fabricated state, the memristor is in the HRS and the negative read bias causes carrier accumulation at the oxide-semiconductor interfaces (Fig. 1bi). The effective index of the optical mode in the Si waveguide underneath therefore changes by the free carrier dispersion effect which in turn leads to an optical phase shift. Setting the memristor to the LRS leads to the formation of CFs (Fig. 1aii) and hence the read voltage can no longer cause carrier accumulation due to the charge leakage (Fig. 1bii), and the free carrier induced phase shift is suppressed.

Fig. 2c shows the reversible switching of a 10 μm radius microring resonator between LRS and HRS states for three consecutive cycles using the device structure shown in Figs. 2a and 2b. Pulses are used instead to switch the memristors, allowing faster operation[28] compared to the I-V sweep. The shaded regions of the spectra indicate the standard deviation between the three switching cycles, clearly revealing the excellent cycle-to-cycle reproducibility. A constant read bias of -3V and a current compliance of 100 nA is applied when taking all the optical measurements in this work unless otherwise stated. The resonances at zero read bias for both the SET and RESET states are shown in Fig. S1 supplementary information. In the LRS, the MOS capacitor is off and -3V read bias only induces a minimal blue shift (see Fig. S1). As the memristor is reset into the HRS, the MOS capacitor is turned on and a blue shift of 0.44nm is observed under the constant -3V read bias, corresponding to a phase shift of 0.09$\pi$ for a phase shifter length of 47μm. The $L_\pi$ is hence estimated to be around 0.5mm - half the length of a ferroelectric-based non-volatile phase shifter[5]. The insertion loss introduced by the carrier dispersion is only 0.28dB or 0.006dB/μm, which is comparable to chalcogenide PCMs[4]. We further show in Fig. 2b that the phase tuning is indeed non-volatile by monitoring the resonance wavelengths over 1 hour where the two phase-levels remain stable over the time, with only minimal drift due to the temperature fluctuations. Supplementary information S4 shows that the resonances measured after 24 hours overlap perfectly with the original spectrum for both the SET and RESET states.

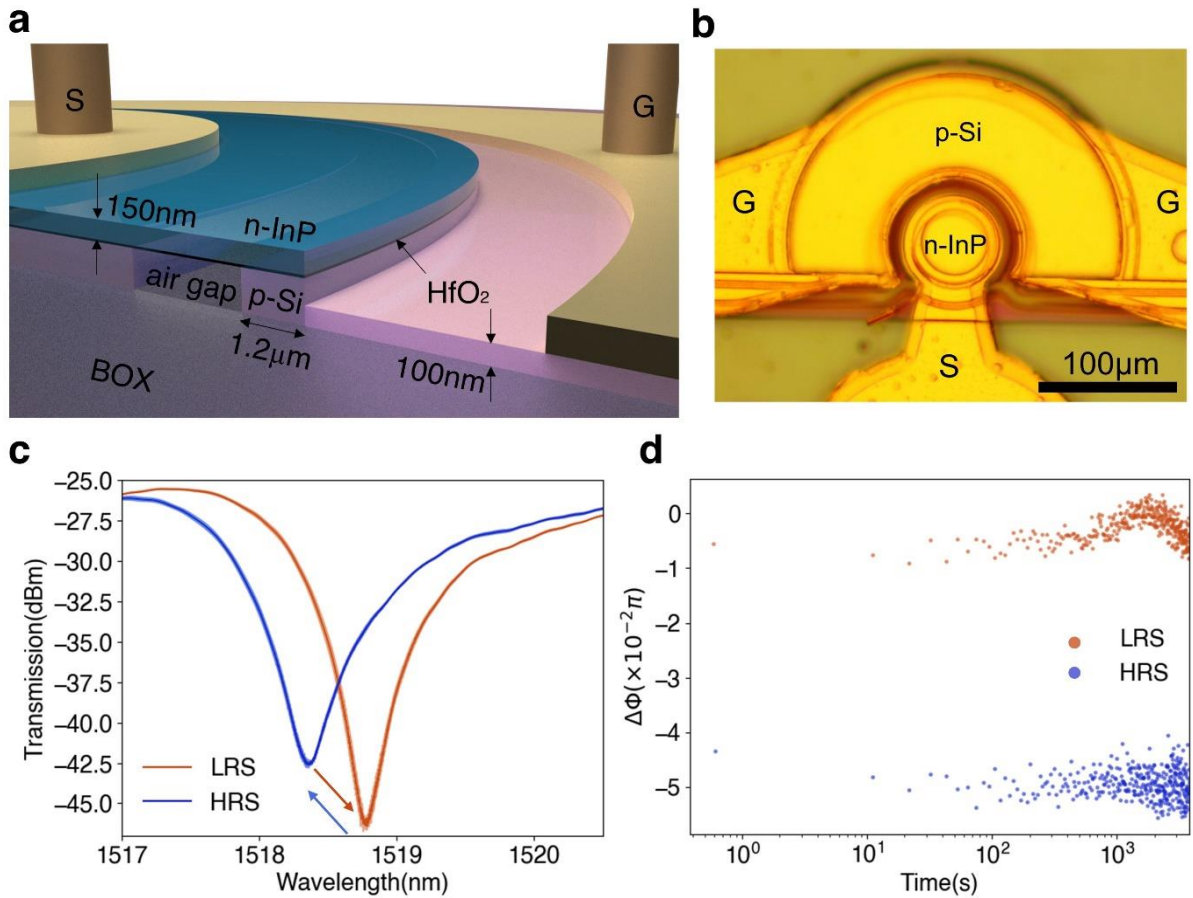

**Figure 2: Non-volatile phase tuning of microring resonator using memristors.**

**a** Schematic of the memristive microring MOS capacitor modulator. BOX stands for buried oxide. S (G), signal (ground) electrode. **b** Optical micrograph of the microring. **c** Reversible switching of microring resonance using memristors. The switching conditions are 7V, 200ns pulse width, 8ns trailing edge for SET and -3V, 200ns pulse width, 8ns trailing edge for RESET. Three consecutive cycles are plotted where the shaded area indicates the standard deviation between the cycles and the solid line indicates the average. **d** Time stability test over 1 hour of the SET and RESET phase states. The optical spectrum is measured every 10 seconds to calculate the resonance wavelength. ΔΦ denotes the optical phase shift. A constant -3V bias at 100nA current compliance is applied to read the optical states in the above measurements.

Device endurance is a key metric in assessing the durability of any non-volatile phase shifter. Here, we performed over 800 cycles or 1,600 transitions on the phase shifter without significant degradation in the performance, see Fig. 3a. The experimental setup and procedure are detailed in Methods. The failure mechanism of the phase shifter is discussed in S5 supplementary information. A phase shift of ~0.03π can still be measured after 800 cycles. Meanwhile in the electrical domain,ced to estimate the resistance. An HRS/LRS ratio of 10× is measured across the 800 cycles where both LHS and RHS exhibit a high resistance of >1GΩ, indicating minimal leakage current (<1nA) under the read bias. The good match between the endurance in optical and electrical domains verifies that the non-volatile phase tuning indeed originates from the memristor switching. Since both states exhibit high resistance (i.e., low switching current ~μA) and the switching time is short (~100ns), we extract a low switching energy of 1.3pJ for SET and 0.4pJ for RESET (see S2 in supplementary information for the estimation of switching energy) by averaging across multiple devices, representing over an order of magnitude reduction in switching energy compared to the state of the art[4,5]. Such low switching energy is an intrinsic advantage of memristor - since the CFs responsible for the switching are only nanometers in scale[29], and only minimal current (~μA) and time (100ns) are required to form and break the filaments. In comparison, PCMs typically require nano-joules to switch due to the micro-scale size[4,30]. The lower RESET energy compared to SET is because the RESET is caused by a compound effect of Joule heating and polarity, as discussed earlier. Because RESET happens at LRS, larger current is drawn at lower voltage compared to HRS. Meanwhile, reversing the polarity causes the retraction of oxygen vacancies, reducing the current required to rupture the filaments. As a result of these two effects, RESET can occur at lower power.

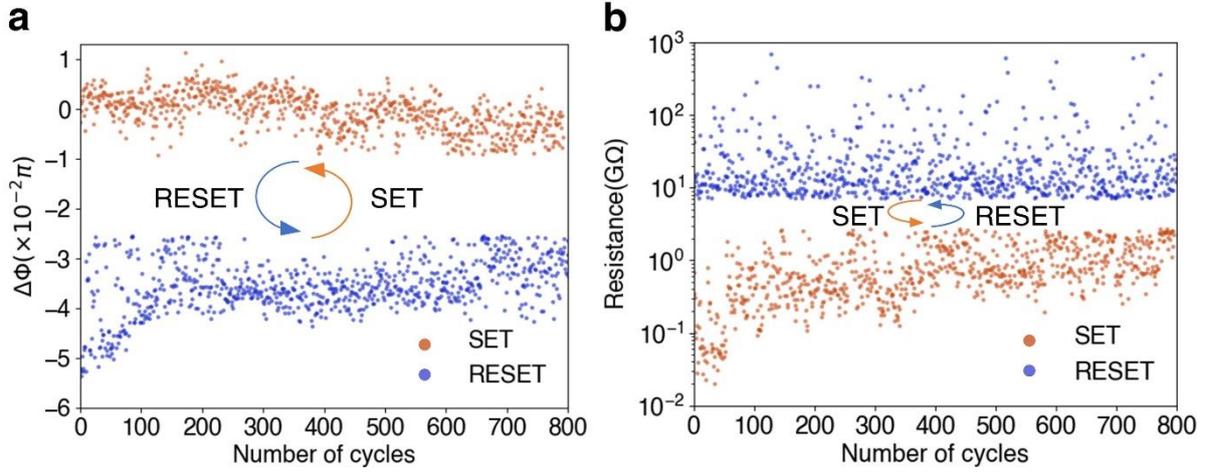

**Figure 3: Endurance of the memristor-based non-volatile phase shifter.**

**a** Cyclability of the phase shifter for 800 consecutive cycles or 1,600 switching events. The switching conditions are 15V, 100ns pulse width, 8ns trailing edge for SET and -3V, 100ns pulse width, 8ns trailing edge for RESET. ΔΦ denotes the optical phase shift. **b** Simultaneous resistance readout of the memristor switching measured at -0.7V read bias over the 800 cycles. The pulse conditions are the same as in **a**.

Finally, we show that the SET and RESET have sub-microsecond optical response time, see Fig. 4. The experimental setup for the real time measurement can be found in Methods. Note that a constant read bias of -2V is applied to the signal during the switching to make sure there is a measurable contrast from the non-volatile switching. Notably, a single 100ns pulse is enough to trigger both the SET and RESET, compared to hundreds of microseconds required for PCMs[4,30] and milliseconds for ferroelectric materials[5]. We estimate an optical response time of ~500ns for SET (Fig. 4a) and ~100ns for RESET (Fig. 4b) respectively. The different response times and dynamics come from the distinct physical effects responsible for SET and RESET in a memristor. When setting the memristor, carrier depletion happens first due to the reverse bias (see Fig. 1bii) and then carrier injection quickly follows once the CFs form. However, since the MOS capacitor is already in the carrier depleted state at zero bias, further carrier depletion hardly gives any optical change. In fact, the slower 'spike' optical response and long relaxation time observed in Fig. 4b is caused by the carrier injection along with the slow thermo-optic effect after the formation of CFs, which is a typical signature of a PIN diode[31] at high current injection. The longer relaxation time indicates that the thermo-optic effect is more dominant. The non-volatile switching is visualized by a permanent change in optical transmission after the transient effect dies out, indicated by the black dash line. On the other hand, when resetting the memristor the optical response changes almost simultaneously with the voltage pulse without any delay or relaxation due the rapid carrier accumulation. The fast response is a signature of carrier accumulation effect where GHz speed has been measured in MOS capacitor modulator[32]. The change of dynamics from a typical carrier injection to carrier accumulation response clearly shows that the non-volatile optical switching is caused by the memristor. It is worth noting that the filament formation in fact occurs at the onset of the voltage pulse (Fig. 4a), implying that sub-nanosecond switching is possible which has been reported in electronic memristors[33].

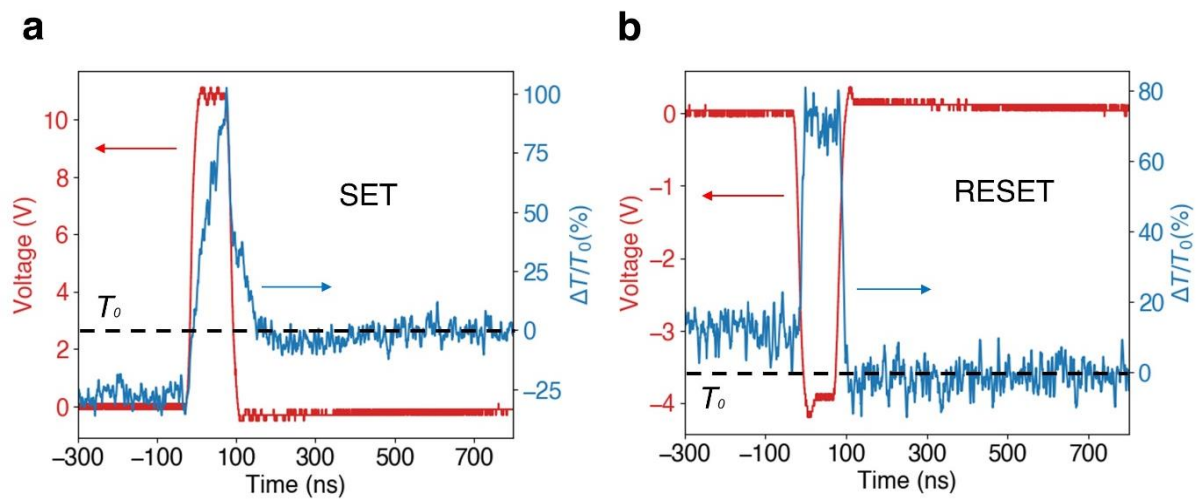

**Figure 4: Real time optical switching response of the phase shifter.**

**a** Dynamics of the RESET operation together with the voltage pulse that triggers the switching. T is the transmission at the microring through port. The black dash line indicates the optical state after transient effect dies out. The pulse conditions are -4.5V, 100ns pulse width, 8ns trailing edge. **b** Dynamics of the SET operation and the voltage pulse that triggers the switching. The pulse conditions are 11V, 100ns pulse width, 8ns trailing edge. A constant -2V read bias is applied in both SET and RESET which is not shown in the figure.

## Conclusion

To summarize, we have demonstrated a non-volatile III-V-on-Si phase shifter with ultra-low switching energy (~400 fJ) and fast response time (~ 100ns). The non-volatile phase shifter also has an excellent endurance of over 800 cycles or 1,600 transitions. The superior performance is made possible by the well-studied electronic memristive effects[17] that have long been explored for non-volatile storage[34] and in-memory computing[35]. Our results show that memristor can be an energy efficient, fast, and reliable technology to realize non-volatile phase tuning in PICs. We believe the device performance can be further improved by adding a transistor to implement fast current compliance during pulsing, also known as the 1T1M (1-Transistor-1-Memristor) configurations[36]. The transistor can prevent current overshoot during the high voltage SET pulse, which is essential to reduce the possibility of device failure (see S4 in Supplementary information). Additionally, by controlling the bias applied to the gate of the transistor, different current limits can be implemented which makes analog operation possible[37]. One can also achieve linear analog operation by adding a $TaO_x$ electro-thermal modulation layer in contact with the $HfO_2$ which can decrease the rate of change of the electric field in the CF gap region[38].

## Methods

*Device Fabrication:* The fabrication[24,39] starts with a 100 mm-diameter SOI substrate with 250 nm-thick, lightly p-doped ($1\times10^{16}$ cm$^{-3}$) top Si layer and 1 μm-thick buried oxide layer (BOX). First, a photolithography step defines the heavily doped p++ regions in Si to form ohmic contact with the metal. We target $1\times10^{20}$ cm$^{-3}$ Boron doping level after ion implantation process and doping activation annealing at 1050°C. We then pattern and dry etch through the top Si layer to form the ring-shaped air trench for electrical and optical isolation along with vertical outgassing channels[23] that ensure proper bonding. To produce the interface HfO$_2$ for bonding, we introduced a plasma activation step in dielectric deposition[40]. Upon depositing ~4.8 nm HfO$_2$ on III-V and Si samples in ALD, III-V/HfO$_2$ and Si/HfO$_2$ samples were manually bonded together to form the heterogeneous MOS capacitor structure, followed by a 300°C annealing step. 250nm SiO$_2$ is then deposited to form a hard mask. Microring patterns of 20 μm diameters are then defined by DUV photolithography and transferred down to the SiO$_2$ hard mask layer, through III-V layers and finally into the Si layer by a self-aligned sequential dry etch process. This step with previously patterned air trench in Si also results in a 1.2 μm wide Si waveguide underneath the InP (Fig. 1a). Another run of photolithography and III-V dry etch are used to remove all III-V material outside the microrings. Metallization to form the terminals on n-InP and p++-Si is followed by metal lift-off process and a rapid thermal anneal at 360°C for 30 seconds. Finally, the wafer is encapsulated by a 300 nm-thick SiO$_2$, followed by etching contact vias to all three terminals and form thick metal probes to conclude the fabrication.

*Experimental setup and measurements:* The wafer was placed on a thermally stabilized stage (at 20°C) and optically probed with a vertical fiber setup via grating couplers defined on the wafer. The devices are electrically addressed via GSG RF probes (CascadeMicrotech ACP-40). The optical spectrum is taken by sweeping the input tunable laser (Santec TSL-510) and collecting at the output using a photodetector (Newport 884-FC) and power meter (Newport 2936-R). The SET and RESET pulses, both with 100 ns pulse width and 8 ns trailing edge, are applied using the HVPGU (high voltage semiconductor pulse generation unit) in the B1500A semiconductor analyzer at the maximum 1 MΩ load impedance. After applying the pulses, the current is read with the SMU (source measurement unit) of B1500A at a bias of -0.7V to estimate the resistance while the optical spectrum is taken using the TSL and power meter. An automated script is used to find the resonance wavelength which corresponds to transmission minima from the spectrum data. The setup schematics are shown in supplementary information S6.

To perform the real-time optical response measurement, the phase shifter is biased at -2V using the SMU so that carrier accumulation is induced inside the device (see S6 in supplementary information). The SET and RESET pulses are applied via the HVSPGU. The output from the grating couplers is amplified by an EDFA and filtered by a wavelength filter. The optical signal from the wavelength filter is detected by a 70GHz high speed photodiode (MACOM P-70A) and measured by a Tektronix 8 GHz real-time oscilloscope.

## Acknowledgement

The author would like to thank Dr. Catherine Graves and Dr. Giacomo Pedretti at HPE for their useful discussion on improving the endurance of the memristors. The author would like to thank the research group of Prof. John E. Bowers at UCSB for sharing testing instruments. Z.F. and A.M. are supported by ONR-YIP grant.

## Author contributions

B.T. and D.L. conceived the projects. Z.F. led the electrical and optical measurements and wrote the script for the endurance test. B.T. prepared the setup for the experiment and helped with the measurements. A.D. helped with the tool automation and performed wafer-scale device screening. D.L. and X.H. designed and fabricated the memristor-based phase shifters. G.Z. took optical microscope images for the devices. B.T., G.Z., A.M. supervised the overall progress of the projects. Z.F. wrote the manuscript with input from all the authors.